\documentstyle[aps,multicol,psfig,eqsecnum]{revtex}
\newcommand{\Iimag}{\tilde{I}}
\newcommand{\omd}{\omega_{\delta}}
\newcommand{\tun}{t}
\newcommand{\Tun}{t}
\newcommand{\curA}{{\cal A}}
\newcommand{\tc}{{T_{\rm c}}}
\newcommand{\xf}{\xi_{\phi}}
\newcommand{\phdag}{{\phantom{\dagger}}}
\newcommand{\ps}{{\phantom{*}}}

\newcommand{\kf}{k_{\rm F}}
\newcommand{\ef}{\epsilon_{\rm F}}
\newcommand{\xa}{\xi_{\rm a}}  
\newcommand{\kb}{k_{\rm B}}
\input{epsf}
\begin{document}
\draft
\title{Andreev interferometry as a probe of 
superconducting phase correlations\\
in the pseudogap regime of the cuprates}
\author{Daniel E.~Sheehy\rlap,\cite{REF:DES} 
Paul M.~Goldbart\rlap,\cite{REF:PMG} 
J\"org Schmalian\cite{REF:JS} and Ali Yazdani\cite{REF:AY}}
\address{
Department of Physics and Materials Research Laboratory, \\
University of Illinois at Urbana-Champaign,
Urbana, Illinois 61801, USA}
  \date{January 20, 2000}
\maketitle
\begin{abstract}
Andreev interferometry---the sensitivity of the tunneling current 
to spatial variations in the local superconducting order at an 
interface---is proposed as a probe of the spatial structure of the 
phase correlations in the pseudogap state of the cuprate 
superconductors.  To demonstrate this idea theoretically, a simple 
tunneling model is considered, via which the tunneling current is 
related to the equilibrium phase-phase correlator in the pseudogap 
state.  These considerations suggest that measurement of the 
low-voltage conductance through mesoscopic contacts of varying areas 
provides a scheme for accessing phase-phase correlation information.  
For illustrative purposes, quantitative predictions are made for a 
model of the pseudogap state in which the phase (but not the amplitude) 
of the superconducting order varies randomly, and does so with 
correlations consistent with certain proposed pictures of the 
pseudogap state. 
\end{abstract}
%
%
%
%
%
\begin{multicols}{2}
\narrowtext
 \section{ Introduction}
\label{sec:intro}
A range of experimental investigations have indicated that underdoped
high-temp{\-}erature superconductors (HTSCs) exhibit intriguing
properties at temperatures {\it above} the superconducting transition
temperature $\tc$.  Most notably, these materials show a strong
suppression in the single-particle electronic spectral weight at low
energies, even at temperatures far above
$\tc$~\cite{AGL96,HD96,CROF98}, a property referred to as the {\em
pseudogap\/}.  A number of scenarios have been proposed to account for
this loss of spectral
weight~\cite{LN92,PWA97,SPS98,EK95,ML96,ER97,SGB97,CAV98}, several of
which invoke the notion that remnants of superconducting correlations
remain in the non-superconducting
state~\cite{EK95,ML96,ER97,SGB97,CAV98}, i.e., that pairing is 
established locally but that it lacks the long-range coherence 
in {\it phase\/} necessary for true superconductivity.

To make progress with understanding the nature of the pseudogap
regime, having experimental access to the {\it spatial structure} of
the correlated electronic state is likely to be of considerable 
%
%
value~\cite{Janko}.
The aim of the present Paper is to identify one possible scheme, involving
low-voltage mesoscopic conductance measurements, for probing this
structure experimentally, and to describe this scheme within the
context of a simple theoretical model.

The basic idea is this.  Let us adopt as a working hypothesis the
picture of the pseudogap regime in which superconductivity is
established locally, but in which the presence and motion of vortices
in the superconducting order parameter cause the
phase of the superconducting order parameter to be randomized beyond
certain correlation length- and time-scales~\cite{REF:avoid}.
The effects of such phase fluctuations on the single-particle
properties of underdoped cuprates have been explored in
Refs.~\cite{FM98,KD98}.
Now, the low-voltage conductance of a normal--to--superconducting
junction includes contributions associated with the Andreev reflection
of quasiparticles from the superconducting
condensate~\cite{REF:conduct}.  What about the low-voltage conductance
of a normal--to--pseudogap junction?  Given the picture of the pseudogap
regime outlined above, and assuming that tunneling through the
junction occurs on a time-scale faster than the time-scale for vortex
rearrangement, we anticipate that there will be contributions to the
conductance due to the Andreev reflection of quasiparticles from the
{\it local\/} superconductivity.  However, owing to the phase
sensitivity of the Andreev reflection process~\cite{REF:Spivak},
any spatial variation in the phase of the superconducting order parameter
over the junction would tend to cause (diffraction-like) interference
of the quasiparticle/hole waves that have been
 Andreev reflected from the junction, and
thereby diminish the associated contribution to the conductance.

\begin{figure}[hbt] 
\epsfxsize=7.0truecm
 \centerline{\epsfbox{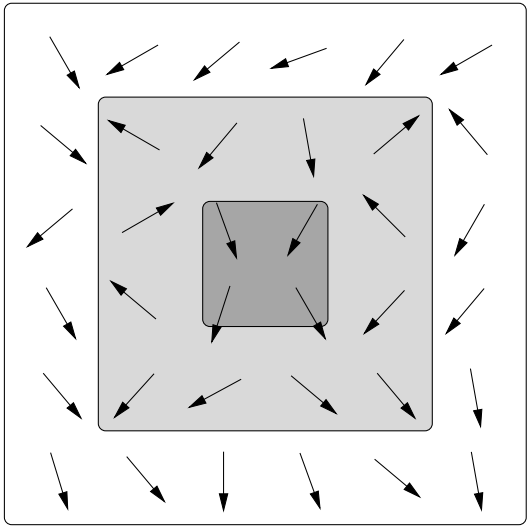}}
\vskip+.5truecm
\caption{Schematic depiction of an instantaneous configuration 
of the superconducting phase in the pseudogap state (arrows).  
The two shades of gray indicate two possible (normal-state) 
contact areas on the pseudogap (i.e.~white) substrate.  Whereas 
the smaller (i.e.~darker) contact abuts a region of nearly 
uniform phase, the larger (i.e.~lighter) contact exhibits 
regions of considerably differing phase.}
\label{FIG:pseudogappict} 
\end{figure}

 Now suppose that the normal contact in a normal--to--pseudogap junction
has a characteristic linear dimension $L$.  If $L$ is smaller than the
characteristic phase-phase correlation (i.e.~inter-vortex) length $\xf$
(e.g.~the smaller contact in Fig.~\ref{FIG:pseudogappict})
then, at any instant, rather little phase variation would be expected
over the contact and the Andreev contribution to the conductance
should be barely diminished.  However, if $L$ is substantially larger
than $\xf$ (e.g.~the larger contact in
Fig.~\ref{FIG:pseudogappict}) then considerable phase variation is
expected over the contact, and the Andreev contribution to the
conductance is likely to be strongly suppressed.  Measurements made
using a range of mesoscopic contact sizes thus have the capability of 
providing a
direct probe of the spatial correlations of the phase of the
superconducting order parameter at various temperatures within the
pseudogap regime.
 
Let us emphasize that the concept of Andreev interferometry is by no
means new; indeed, several groups have considered this concept both
theoretically~\cite{REF:Hekking} and experimentally~\cite{REF:Andint}.
However, to the best of our knowledge this interferometry has
primarily been considered in contexts in which reflection is from a
truly superconducting region (rather than from a pseudogap region),
and in settings in which the phase has an average value that varies
in a relatively simple way in space (such as on either side of a Josephson
junction).  Here, we are considering a setting in which the
interferometry is being used as a probe of the superconducting
 fluctuations.
It should be mentioned that in recent work Choi et
al.~\cite{REF:Choi} considered the issue of whether or not the
zero-bias tunneling conductance peak~\cite{REF:Covington} would
survive at temperatures above $\tc$.  This work involves applying the
BTK technique~\cite{REF:BTK} to the physical picture of the pseudogap
regime explored, e.g., in Refs.~\cite{FM98,KD98}.  It amounts to a
computation of the conductance of a normal--to--superconductor interface
for a d-wave superconductor in a uniform supercurrent-carrying state,
this conductance then being averaged over a Gaussian distribution of
uniform supercurrents (in order to model the pseudogap state).  This
yields a conductance dependent upon the statistical distribution of
local values of the supercurrent arising from varying vortex
locations.  In contrast, the present work focuses on the spatial 
correlations of the phase in the pseudogap regime and, specifically, 
how such correlations may be accessed experimentally. 

\section{Tunneling current for a Normal--To--Pseudogap junction}
\label{sec:tunneling current}
We now illustrate the ideas of Sec.~\ref{sec:intro} by computing the
conductance of a normal--to--pseudogap junction within the tunneling
formalism, and show how this conductance depends on the pseudogap
phase-phase correlation function.  To this end, we adopt as the
tunneling Hamiltonian $H_{\rm T}$~\cite{Mahan} between a normal state
(N) and a pseudogap state (P):
\begin{equation}
H_{\rm T} \equiv \sum_{\sigma = \pm} \! \int_{\rm P \/} 
d^3r \!
                                     \! \int_{\rm N \/}
d^3s
          \left(  \Tun^{\ps}_{{\bf r},{\bf s}}  \,
                    d^{\dagger}_{{\bf r},\sigma} \,
                    c^{\phdag}_{{\bf s},\sigma} 
           +      \Tun^*_{{\bf r},{\bf s}}  \,
                    c^{\dagger}_{{\bf s},\sigma} \,
                    d^{\phdag}_{{\bf r},\sigma} \right),
\end{equation}
where the position ${\bf s}$ lies on the normal side 
of the junction
and the position ${\bf r}$ lies on the pseudogap side.  
The operators $c^{\phdag}_{{\bf s},\sigma}$ 
(or $c^{\dagger}_{{\bf s},\sigma}$) and 
 $d^{\phdag}_{{\bf r},\sigma}$ 
(or $d^{\dagger}_{{\bf r},\sigma}$) 
respectively annihilate (or create) quasiparticles
with spin projection $ \sigma $ on the normal side 
at ${\bf s}$ and on the pseudogap side at ${\bf r}$.  
We choose the
interface to be in the plane $z = 0$ (where $\{ x,y,z \}$ are
Cartesian coordinates and $\{
{\bf{e}}_{x},{\bf{e}}_{y},{\bf{e}}_{z} \}$ are the
corresponding basis vectors) and, accordingly, decompose vectors such as
${\bf s}$ and ${\bf r}$ into components parallel 
(e.g.~${\bbox{\sigma}}$) and perpendicular 
(e.g.~$s_z {\bf{e}}_{z}$) to
the interface so that ${\bf s} = {\bbox{\sigma}} +
s_{z}{\bf{e}}_{z} $ and ${\bf r} = {\bbox{\rho}} +
r_{z}{\bf{e}}_{z} $.  This choice, together with the assumption
that tunneling only occurs locally at the interface leads us to assert
that the tunneling matrix elements $\tun_{{\bf r},{\bf s}}$ are given by
\begin{equation}
\Tun_{{\bf r},{\bf s}} = \tun_0 a
\,\delta^{(1)}(s_z)\,\delta^{(1)}(r_z)\, \delta^{(2)}({\bbox{\sigma}}
- {\bbox{\rho}}),
\end{equation}
where $a$ is a microscopic length scale characterizing 
the thickness of the ``active'' layer for tunneling of 
particles and $\tun_0$ is the typical energy 
scale for this process.  

Next, we compute the current $I(V)$ as a function of the voltage $V$.  
To do this, we consider the expectation value of the tunneling current 
operator $ [-eQ_{\rm N},H_{\rm T}]/i\hbar$, where $-e$ is the electron 
charge:
\FL
\begin{equation} 
-\frac{ie}{\hbar}\sum_{\sigma = \pm}                
\int_{\rm P \/}
\!\!
d^3r 
\!\!
\int_{\rm N \/}
\!\! 
d^3s
\Big(
 \Tun_{{\bf r},{\bf s}}\,
d^{\dagger}_{{\bf r},\sigma}\,
c^{\phdag}_{ {\bf s},\sigma}
-\Tun_{{\bf r},{\bf s}}^{\ast}\,
c^{\dagger}_{ {\bf s},\sigma}
d^{\phdag}_{{\bf r},\sigma}
\Big),
\end{equation}
with respect to the full Hamiltonian for the system, i.e., 
$H = H_{\rm N}+H_{\rm P}+H_{\rm T}$, where $H_{\rm N/P}$ is the 
Hamiltonian for the normal/pseudogap side and $-eQ_{\rm N}$ is 
the charge operator for the normal side.  In fact, it is 
convenient to obtain $I(V)$ perturbatively in the tunneling 
amplitude $\tun_0 a$ by applying the Matsubara technique to the 
imaginary-time dependent tunneling current 
$\Iimag(\tau_1)$~\cite{Tsuzuki}.  The lowest-order 
term, which is of order $|\tun_0 a|^2$, represents the normal 
(i.e.~single-particle) current.  This contribution is suppressed at 
low voltages due to the presence of a gap at low energies on the 
pseudogap side.  The next-order contribution to $\Iimag$, which is 
of fourth order in $\tun_0 a$, is given by
\begin{eqnarray}
&&\Iimag(\tau_1) = -\frac{e}{\hbar}
 |\tun_0 a|^4
 \! \sum_{{\bf \sigma}_i=\pm} \int_{\curA}
   \prod_{j=1}^4 d^2 \rho_j \, 
\!\int_0^\beta\!\!d\tau_2\,d\tau_3\,d\tau_4\,
                                     \nonumber \\
&&\quad\times
              \big\langle T_{\tau} \,
                     d^{\dagger}_{{\bbox{\rho}}_1, \sigma_1}(\tau_1) \,
                     d^{\dagger}_{{\bbox{\rho}}_2, \sigma_2}(\tau_2) \,
                     d^{\phdag}_{{\bbox{\rho}}_3, \sigma_3}(\tau_3) \,
                     d^{\phdag}_{{\bbox{\rho}}_4, \sigma_4}(\tau_4)
              \big\rangle_{\rm P\/} \nonumber \\
&&\quad\times
              \big\langle  T_{\tau} \,
                     c^{\phdag}_{{\bbox{\rho}}_1,  \sigma_1}(\tau_1) \,
                     c^{\phdag}_{{\bbox{\rho}}_2,  \sigma_2}(\tau_2) \,
                     c^{\dagger}_{{\bbox{\rho}}_3, \sigma_3}(\tau_3) \,
                     c^{\dagger}_{{\bbox{\rho}}_4, \sigma_4}(\tau_4)  
             \big\rangle_{\rm N\/}  \nonumber \\
&&\quad\times \label{eq:jtau}
         \, {\rm e}^{-i\omd(\tau_2+\tau_3+\tau_4)} \,    
          {\rm e}^{-i\Omega(\tau_1+\tau_2-\tau_3- \tau_4)},
                                    \end{eqnarray}
where $\langle \cdots \rangle_{\rm P/N} $ indicates an equilibrium 
expectation value with respect to $H_{\rm P/N}$, and $\beta$ measures 
the inverse temperature.  Operators such as 
$c^{\phdag}_{{\bf r},\sigma}(\tau)$ 
        [or $d^{\phdag}_{{\bf s},\sigma}(\tau)$]
are interaction-picture operators, i.e., 
${\rm e}^{ K_0\tau}
 c^{\phdag}_{{\bf r},\sigma}
 {\rm e}^{-K_0\tau}$ 
        (or 
        ${\rm e}^{ K_0\tau}
         d^{\phdag}_{{\bf s},\sigma}
         {\rm e}^{-K_0\tau}$), 
where 
$K_0\equiv H_{\rm N}-\mu_{\rm N}Q_{\rm N}$ 
        (or $K_0\equiv H_{\rm P}-\mu_{\rm P}Q_{\rm P}$).
Here, $\mu_{\rm N}$ (or $\mu_{\rm P}$) is the chemical potential on the 
normal (or pseudogap) side, and  $Q_{\rm N}$ (or $Q_{\rm P}$) is the 
charge operator for the normal (or pseudogap) side. The physical current 
$I(V)$ is given by the imaginary part of $\Iimag(\tau_1)$ after making 
the following analytical continuations: 
$i\omd\rightarrow i0^{+}$, 
$i\Omega\rightarrow eV$ (i.e.~the voltage across the junction), and  
$-i\hbar\tau_1 \rightarrow t$ (i.e.~the time).  

\begin{figure}[hbt] \epsfxsize=3.0truein
 \centerline{\epsfbox{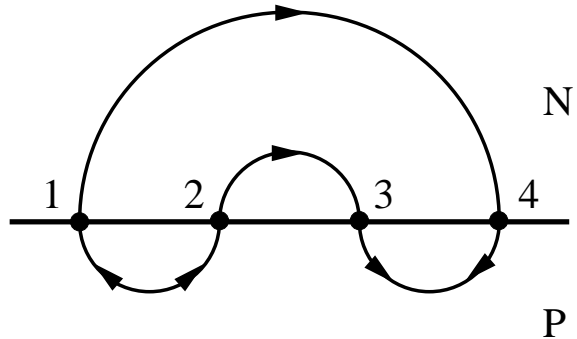}}
\vskip+.3truecm 
\caption{Diagram depicting the leading-order (in tunneling 
amplitude) phase-sensitive contribution to the current.} 
\label{FIG:diagrampict}
\end{figure}

To apply Eq.~(\ref{eq:jtau}) to the setting at hand, namely one side
of the junction being normal and the other being in the pseudogap
regime, we shall need to evaluate the two two-particle Green function
factors that feature in it, one for the normal side and one for the
pseudogap side.  
For the normal side, we assume that the corresponding two-particle 
Green function is factorizable into two single-particle
Green functions, i.e.,
\begin{eqnarray}
       && \langle  T_{\tau} \,c^{\phdag}_{{\bbox{\rho}}_1,  +}(\tau_2) \,
          c^{\phdag}_{{\bbox{\rho}}_2,  -}(\tau_2) \,
                     c^{\dagger}_{{\bbox{\rho}}_3, -}(\tau_3) \,
                     c^{\dagger}_{{\bbox{\rho}}_4, +}(\tau_4)  
             \big\rangle_{\rm N\/}, 
\nonumber \\
&&\qquad 
= 
G^{\rm N}({\bbox{\rho}}_1,{\bbox{\rho}}_4;\tau_1-\tau_4)\,
G^{\rm N}({\bbox{\rho}}_2,{\bbox{\rho}}_3;\tau_2-\tau_3),
\end{eqnarray}
where 
$G^{\rm N}({\bbox{\rho}},{\bbox{\rho}}';\tau - \tau')
\equiv \langle  T_{\tau}\,  c^{\phdag}_{{\bbox{\rho}},  \sigma}(\tau) \,
                        c^{\dagger}_{{\bbox{\rho}}', \sigma}(\tau') 
\rangle $
is the single-particle Green function on the normal side. 
On the other hand, for the pseudogap side we adopt
a model in which the pseudogap state is a 
superconductor
that is ``disordered'' by a {\it static} pattern of vortices and
characterized by a {\it static} phase-phase correlator.
We do not require
information concerning the dynamic phase-phase correlation function 
because we are assuming that the Andreev reflection
process is rapid, compared with the time needed for vortices to
substantially rearrange the phase structure.  To support this
assumption, let us note that the time-scale associated with Andreev
reflection is of order $\tau_{\rm AR} \sim \xa/v_{\rm F}$,
where $v_{\rm F}$ is the Fermi velocity of the incoming electron and
$\xa$ is the amplitude-fluctuation correlation length (i.e.~the
Cooper-pair size) on the pseudogap side. Then, by using the estimates
$v_{\rm F}\approx 10^{7} {\rm cm}/{\rm s}$ and 
$\xa\approx 1\,{\rm nm}$ (typical for a HTSC) we find that 
$\tau_{\rm AR} \sim 10^{-14}\,{\rm s}$.  
The experiments of Corson et al.~\cite{ref:corson} 
indicate that the vortex-pattern rearrangement
time corresponds to frequencies in the terahertz range (i.e.~is of
order $10^{-9}\,{\rm s}$) so that, at least as a starting point, we
may neglect the dynamics of the vortices.  Thus, we assume that
quasiparticles incident from the normal side effectively encounter, and
are Andreev reflected by, a static pair-potential that has a nonzero
amplitude (except at the vortex cores, which are small) and a
spatially random phase.  With this in mind, we characterize the pseudogap
side by the anomalous Green function
\begin{eqnarray}
\label{eq:pggf}
F^{\rm P}({\bf r},{\bf r}';\tau,\tau')& \equiv&
\langle T_{\tau} \, d_{{\bf r}, \downarrow}(\tau)
               \,  d_{{\bf r}', \uparrow}(\tau')\rangle_{\rm P}
\\
&=&
 f_0({\bf r} - {\bf r}'; \tau -\tau')\, {\rm e}^{i(\phi({\bf r}) 
+ \phi({\bf r}'))/2}, \nonumber 
\end{eqnarray}
where the phase $\phi({\bf r})$ varies randomly in space, and the
function $f_0$ is given by the value it takes in a conventional
superconductor~\cite{REF:AGD}, i.e.,
\begin{eqnarray}
f_0({\bf r};\tau)
&=& \int \frac{d^3 k}{(2\pi\hbar)^3} 
\frac{1}{\beta}\sum_{n=-\infty}^{\infty} 
        {\rm e}^{i\omega_n\tau}
        {\rm e}^{i{\bf k}\cdot{\bf r}/\hbar}
\nonumber 
\\
&& \times \left(\frac{\Delta_{\bf k}}{2E_{\bf k}}
          \right) 
          \left(
                         \frac{1}{i\omega_n + E_{\bf k}}
                    -    \frac{1}{i\omega_n - E_{\bf k}}
                                     \right). 
\end{eqnarray}
Here, $E_{\bf k}$ [$\equiv\sqrt{\xi_{\bf k}^2 + \Delta_{\bf k}^2}$, 
with $\xi_{\bf k} \equiv (\hbar^2 k^2/2m)-\mu$] is
the excitation energy in the pseudogap material and $\Delta_{\bf k}$ is
the gap amplitude.
The Matsubara frequencies $\omega_n$ are defined to be $\omega_n =
(2n+1)\pi/\beta $ for integer $n$.
For the sake of simplicity, we now focus on the case of 
s-wave pairing and thus set $\Delta_{\bf k} = \Delta$~\cite{d-wave}.
We approximate the two-particle Green function on the ${\rm P}$ side 
in Eq.~(\ref{eq:jtau}) by making a Gorkov factorization into the 
anomalous Green function $F^{\rm P}$ and its conjugate.  In principle, 
there will also be a contribution associated with factorization into 
normal Green functions.  However, these contributions are suppressed 
at low voltages, and thus it is adequate to take for the current
\begin{eqnarray}
\Iimag(\tau_1)
 &=& 
 -\frac{2e}{\beta}  |\tun_0 a|^4\!
\int\limits_{\curA} \prod_{j=1}^4 \!d^2\!{\rho}_j 
                {\rm e}^{i(\phi({\bbox{\rho}}_1)+\phi({\bbox{\rho}}_2)-
                    \phi({\bbox{\rho}}_3)-\phi({\bbox{\rho}}_4))/2} 
\nonumber \\
&& \times 
\sum_{n=-\infty}^{\infty}
G^{\rm N}({\bbox{\rho}}_2,{\bbox{\rho}}_3;i\omega_n +i\omd +i\Omega) 
 \nonumber
\\ 
&& \,\times 
G^{\rm N}({\bbox{\rho}}_1,{\bbox{\rho}}_4;-i\omega_n -3i\omd +i\Omega)
 \nonumber
\\
&&      
\times \,f_0({\bbox{\rho}}_1-{\bbox{\rho}}_2;i\omega_n) \,
       \,f_0({\bbox{\rho}}_3-{\bbox{\rho}}_4;i\omega_n+2i\omd),
      \label{eq:bigone} 
\end{eqnarray}
where the limit $\curA$ indicates that the interface integrals are
constrained
 to the area of the contact between the ${\rm N}$ and ${\rm P}$ regions.
Equation~(\ref{eq:bigone}) may be
expressed diagrammatically, as shown in Fig.~\ref{FIG:diagrampict},
where one-arrow lines denote normal Green functions,  two-arrow
lines denote anomalous Green functions (and the straight line denotes the
interface).  We see that this contribution to the current involves the
correlation of an electron and a hole propagating on the normal side,
mediated by the static random pair-potential on the pseudogap side of
the junction.

Equation~(\ref{eq:bigone}), which represents the 
leading contribution due to Andreev reflection at an interface, 
may be considerably simplified in situations in which 
$\xa \ll \xf$ (i.e.~the phase order which
we are interested in probing via Andreev reflection persists over
length scales much larger than the pair size), which is not only
the case for the usual NS interface setting but also for 
the present NP setting.  To support this
assertion, we obtain the ratio of these two length-scales by examining
the results of the Corson et al. experiments on the 
high-frequency AC conductivity of 
${\rm Bi}_2
 {\rm Sr}_2
 {\rm Ca}
 {\rm Cu}_2
 {\rm O }_{8+\delta}$~\cite{ref:corson}.
According to the analysis of  Corson et al., 
in the pseudogap state the ratio of
$\xf$ to $ \xa$ is related to the vortex diffusion time $\tau$
via
\begin{equation}
\left(\frac{\xf}{\xa}\right)^2 = \frac{\tau \Omega_0}{2\pi},
\end{equation}
where $\Omega_0$ is a parameter determined by Corson et al. to be
$1.14\times 10^{14}\, {\rm s}^{-1}$.  From Fig.~(4) of 
Corson et al., we see
that at $T = 75\,{\rm  K}$, $\tau \sim 10^{-12} \,{\rm s}$, 
so that $\xf^2/\xa^2 \sim 20$.

The significance of the separation of the length-scales $\xa$ 
and $\xf$ in 
the present context follows from the fact that the function $f_0$ has spatial 
range $\xa$.  Thus, in Eq.~(\ref{eq:bigone}) the spatial 
integrations over 
the coordinates $\{ {\bbox{\rho}}_i \}_{i=1}^4$ may be simplified because 
$f_0$ varies rather more rapidly in space than do the other factors in the 
integrand.  This allows us to make the approximation
\begin{eqnarray} 
f_0({\bbox{\rho}},i\omega) &\approx & \delta^{(2)}({\bbox{\rho}})\, 
                    \int_{\curA}d^2{\bbox{\rho}'}  \, 
                         f_0({\bbox{\rho}'},i\omega) 
\nonumber \\
 &\approx &  \delta^{(2)}({\bbox{\rho}})\,\int_{-\infty}^{\infty} 
                  \frac{dk_{z}}{2\pi\hbar} \,    
              \hat{f}_0(k_{z}{\bf{e}}_{z}, i\omega), \label{eq:fnought}
\end{eqnarray}
where $\hat{f}_0$ is the (three-dimensional) Fourier transform
of $f_0$ and 
$k_{z}$ is the momentum component perpendicular to the interface.
By using this approximation, we obtain  
\begin{eqnarray} 
&&\Iimag(\tau_1)
 =  -\frac{2e}{\beta}|\tun_0 a|^4 \, 
 \sum_{n=-\infty}^{\infty}
\int_{\curA} d^2\rho_1 \, \int_{\curA}d^2\rho_2 \,
          {\rm e}^{i(\phi({\bbox{\rho}}_1)-
                    \phi({\bbox{\rho}}_2))} 
\nonumber \\
&& \qquad\times   
G^{\rm N}({\bbox{\rho}}_1,{\bbox{\rho}}_2;i\omega_n + i\omd + i\Omega) 
\nonumber 
\\ 
&& \qquad\times 
G^{\rm N}({\bbox{\rho}}_1,{\bbox{\rho}}_2;-i\omega_n -3i\omd +i\Omega)
 \label{eq:bigone2}
\\
&& \qquad    \times  \int   \frac{dk_{z}}{2\pi\hbar} \,
     \hat{f}_0(k_{z}{\bf{e}}_{z}, i\omega_n)\,
  \int   \frac{dk_{z}'}{2\pi\hbar} \, 
      \hat{f}_0(k_{z}'{\bf{e}}_{z}, i\omega_n+2i\omd). \nonumber
\end{eqnarray}
Having derived 
 Eq.~(\ref{eq:bigone2}), an equation applicable to any given 
realization of the phase field  $\phi({\bbox{\rho}})$ on the 
${\rm P}$-side of the interface, we conclude the present section 
by performing the  averaging this current over an 
as-yet-unspecified distribution 
of phase fields.  As discussed above, the time-scale for the 
tunneling process is shorter than the time-scale for phase 
rearrangement.  Thus, it is appropriate to proceed as we have, 
by first computing the current for a fixed realization of the 
phase field, and then to construct the time-averaged current, 
averaged over times longer than the phase rearrangement time, 
by averaging the current over an appropriate (in this case, 
equilibrium) distribution of phase fields.  
Denoting such averaging by $[\cdots]$, and introducing the appropriate 
phase-phase correlator
\begin{equation}
g({\bbox{\rho}}_{1}-{\bbox{\rho}}_{2}) 
\equiv 
\big[
{\rm e}^{ i\phi({\bbox{\rho}}_{1})}\,
{\rm e}^{-i\phi({\bbox{\rho}}_{2})}
\big],
\end{equation}
we arrive at a formula for the time-averaged current 
$\big[ \Iimag(\tau_1)\big]$, i.e., Eq.~(\ref{eq:bigone2}) 
but with the phase factors 
$\exp i\big(\phi({\bbox{\rho}}_1)-\phi({\bbox{\rho}}_2)\big)$
replaced by $g({\bbox{\rho}}_{1}-{\bbox{\rho}}_{2})$. 
For convenience, we express the normal-side Green function in terms the
corresponding spectral function $A$:
\begin{equation}
G^{\rm N}({\bbox{\rho}}_1,{\bbox{\rho}}_2;i\omega_n)
\equiv \int_{-\infty}^{\infty} \frac{d\epsilon}{2\pi} 
\frac{A({\bbox{\rho}}_1,{\bbox{\rho}}_2;\epsilon)}{i\omega_n - \epsilon}.
\end{equation}
Then we may perform the integrations over $k_z$ and 
$k_z'$ 
(by converting them to energy integrals), 
as well as the summation over Matsubara
frequencies.  By performing the necessary analytic continuations
and taking the imaginary part, we obtain an expression for the 
tunneling current $I(V)$ through a mesoscopic interface
between a normal metal and a material in the pseudogap state:
\begin{eqnarray}
I(V) &=& 
\frac{e}{\hbar}\frac{\pi}{8}|\tun_0 a|^4
     \tilde{\nu}_{\rm P}^2 
\int_{\curA} d^2\rho_1  \int_{\curA} \! d^2\rho_2  \,\, 
          g({\bbox{\rho}}_1-
                    {\bbox{\rho}}_2) 
\nonumber \\
&&\times \int_{-\mu}^{2eV+\mu} \, d\epsilon  
\frac{\Delta^2}{\Delta^2-(eV-\epsilon)^2}
\big\{n(\epsilon-2eV)-n(\epsilon)\big\}
\nonumber \\
&& \times 
A({\bbox{\rho}}_1,{\bbox{\rho}}_2;\epsilon) \,
A({\bbox{\rho}}_1,{\bbox{\rho}}_2;2eV-\epsilon).
\label{final}
\end{eqnarray}
where $n(\epsilon) \equiv (\exp(\beta \epsilon)+1)^{-1}$ is the Fermi
distribution function and $\tilde{\nu}_{\rm P}$ 
[$\equiv  m/(2\pi\hbar^2\kf)^{-1}$
with $\kf$ being the Fermi wave-vector] is 
the one-dimensional density of 
states on the P-side.
\section{Case of clean normal-metal contact}
\label{sec:clean}
\subsection{General considerations}
\label{sec:clean_intro}
In this section we pursue the evaluation  of Eq.~(\ref{final}) for the 
case of a normal contact that is perfectly clean, in the sense
that  the spectral 
function $A^{\rm C}$ (with the superscript
C standing for clean) has the form appropriate for a pure metal:
\begin{equation}
A^{\rm C}({\bf p};\epsilon) 
= 2\pi\,\delta^{(3)}(\epsilon_p - \epsilon),
\end{equation}
in which $\epsilon_p \equiv p^2/2m - \mu$.
The (three-dimensional) Fourier transform of this quantity is given by
\begin{mathletters}
\begin{eqnarray}
A^{\rm C}({\bf r},{\bf r}';\epsilon) 
&\equiv& \int \frac{d^3p}{(2\pi\hbar)^3} \,
A^{\rm C}({\bf p};\epsilon)\, {\rm e}^{i{\bf p}\cdot ({\bf r} -{\bf r}')/\hbar},
\\
&=& 
\frac{m}{\hbar^2\pi}\frac
{\sin\big\{\sqrt{2m\hbar^{-2}(\epsilon+\mu)}|{\bf r}-{\bf r}'|\big\}}
{|{\bf r}-{\bf r}'|},
\label{cleanspectral}   
\end{eqnarray}
\end{mathletters}
By inserting this expression into Eq.~(\ref{final}), and limiting our 
attention to low temperatures (i.e.~$\kb T \ll eV$, with $\kb$ being 
Boltzmann's constant) and low voltages 
(i.e.~$eV \ll \Delta$)~\cite{REF:inequals}, 
we obtain an equation for the low-voltage conductance as a functional 
of the pseudogap phase-phase correlation function
\begin{eqnarray}
\label{eq:fin}
&&\left.\frac{I(V)}{V}\right|_{V\rightarrow 0^{+}} = 
\frac{e^2}{\hbar}
\left| \frac{\tun_0\kf a}{4\pi\ef}\right|^4
 \kf^2\pi
\\ 
&& \quad
\times\int_{\curA} d^2\rho_1  
\int_{\curA}\!
d^2\rho_2 \,\,  
          g({\bbox{\rho}}_1-
                    {\bbox{\rho}}_2)
\frac{\sin^2 k_{\rm F}|{\bbox{\rho}}_1-{\bbox{\rho}}_2|}
{|{\bbox{\rho}}_1-{\bbox{\rho}}_2|^2}. \nonumber 
\end{eqnarray}
 \subsection{Illustrative example: BKT correlations}
\label{sec:example}
The main conclusion of Sec.~\ref{sec:clean_intro} is that the contribution
of Andreev reflection to the tunneling current is sensitive to 
spatial inhomogeneity of the superconducting phase, such as is
proposed to exist in the pseudogap state.
For the purposes of illustration, we now examine a specific example of
how the current enhancement due to Andreev reflection
is increasingly
 suppressed, with increasing area, due to destructive interference.
In this example, we assume that the phase-phase correlations 
in the pseudogap state are adequately modeled by those 
associated with the
Berezinski\i-Kosterlitz-Thouless (BKT) theory
of the two-dimensional $XY$ model~\cite{Ber70,KT73,Kosterlitz}. 
The relevance of this theory to the cuprate materials~\cite{REF:exEK95}
originates in the fact that their pronounced planar character 
causes the intermediate length-scale electronic structure
 to be characterized 
by two-dimensional $XY$ behavior, which 
is expected to 
cross over to 
three-dimensional $XY$ behavior only very close to the transition.  
In order to compute the current for this BKT scenario, we need a form for 
$g({\bbox{\rho}})$.  
On length-scales short compared with the phase-phase correlation length 
$\xf$, the function
$g({\bbox{\rho}})$ approaches unity; on length-scales long compared
with $\xf$,  it
exhibits exponential behavior~\cite{Kosterlitz}.  
As we are only seeking  an illustrative computation of the
current, 
the exact details of this 
crossover are unimportant,  
 and thus we adopt the form 
\begin{equation}
g({\bbox{\rho}})= 
{\rm e}^{-|{\bbox{\rho}}|/\xf},
\label{eq:g(r)}
\end{equation}
and we take the interface to have  the shape of a disk of radius $L$.
Inserting Eq.~(\ref{eq:g(r)}) into Eq.~(\ref{eq:fin}),
we see  that the low-voltage Andreev conductance per unit area
through the interface has the form 
\begin{mathletters}
\begin{eqnarray}
&&\left.\frac{I(V)}{\pi L^2\,V}\right|_{V\rightarrow 0^{+}} 
= \Gamma_{\rm C} \,f_{\rm C}(\kf \xf,L/\xf),
\\
&&\Gamma_{\rm C} \equiv 
\frac{e^2}{\hbar}
\frac{\kf^2\pi^2}{2}
\left| \frac{\tun_0k_{\rm F}a}{4\pi\epsilon_{\rm F}} \right|^4
   \times \ln\big(1+4\kf^2\xf^2\big),
\\
&& f_{\rm C}(\kf \xf,L/\xf) \equiv
\frac{2}{\pi^2 \ln\big(1+4\kf^2\xf^2\big)} \times
\nonumber
\\
&& \quad
\int_{1} d^2x_1 \, \int_{1} d^2x_2 \,  
\frac{\sin^2 k_{\rm F}L|{\bf x}_1-{\bf x}_2|}
{|{\bf x}_1-{\bf x}_2|^2}
          {\rm e}^{-|{\bf x}_1-
                    {\bf x}_2|(L/\xf)}. \label{eq:fclean}
\end{eqnarray}
\end{mathletters}
Here, the subscript $1$ indicates that the integrals are taken over
disks of unit radius.  The  prefactor $\Gamma_{\rm C}$ 
is the limiting value of the conductance per unit
area in the limit of large interface area.
[Note that $\Gamma_{\rm C}$ vanishes for the case of 
no phase coherence (i.e.~$\kf \xf = 0$).]

One of our primary concerns is how the 
varying of the interface size would provide information
regarding the structure of the phase correlations; this information
is contained in the function $f_{\rm C}$, which 
depends only on the dimensionless quantities $\kf \xf$ and
$L/\xf$.  For 
generic values of its arguments, the form of $f_{\rm C}$ 
can be determined only via numerical integration; however, its
behavior can be determined  
in various physically relevant asymptotic
limits.  
To begin with, 
let us assume that the phase correlations persist over length-scales
that are long compared with the Fermi wavelength on the normal 
side (i.e.~$\kf \xf \gg 1$), and let us consider varying the interface size.
For small interface sizes (i.e.~$ \kf^{-1} \ll  L \ll \xf$),
$f_{\rm C}$ increases logarithmically with $L$ 
(i.e.~$f_{\rm C}\approx \ln \kf L/\ln \kf \xf$).  In the opposite regime
of large interface sizes, we expect that 
Andreev reflection will occur from 
independent ``domains'' of uniform phase  (so that, e.g., the doubling 
of the 
area should double the conductance).
Indeed, for $L \gg \xf$, 
\begin{equation}
f_{\rm C} \approx  1 - 
\frac{16\,\kf^2\,\xf^2}{(1+4\,\kf^2\,\xf^2)\ln(1+4\,\kf^2\,\xf^2)}
\left(\frac{\xf}{\pi L}\right)
\end{equation}
for any value of $\kf \xf$.
\begin{figure}[hbt]
\epsfxsize=8.0cm
 \centerline{\epsfbox{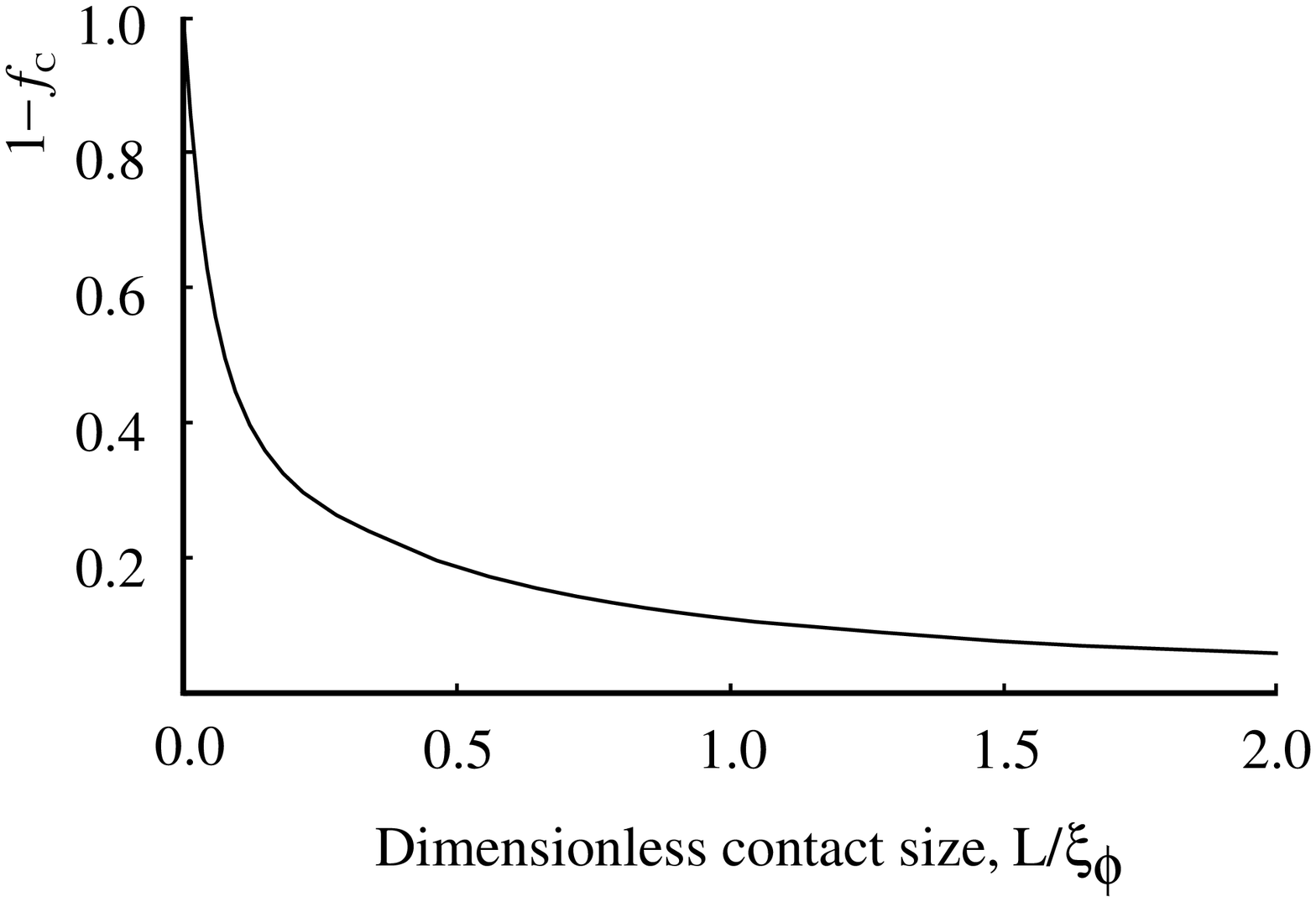}}
\vskip+.2truecm 
\caption{$1-f_{\rm C}$ (i.e.~the departure of the dimensionless 
Andreev conductance per unit area from its large-area limit) 
as a function of the dimensionless size of the interface 
$L/\xf$, for the case of $\kf \xf = 100$ (computed numerically).  
For $L$ much smaller than $\xf$, the zero-voltage conductance 
per unit area is much smaller than its asymptotic value.} 
\label{FIG:fclean} 
\end{figure}

 To study the behavior of $f_{\rm C}$ for intermediate values of 
$L$, we perform the integrals in 
Eq.~(\ref{eq:fclean}) numerically.  In Fig.~\ref{FIG:fclean},
we show
$1 - f_{\rm C}$ as a function of $L/\xf$ for the case of long-range 
phase correlations (i.e.~$\kf\xf = 100$).  
Thus, as discussed Sec.~\ref{sec:intro}, a series of mesoscopic
conductance measurements
involving a range of contact sizes is expected to 
be rather sensitive to the characteristic length-scale 
of phase correlations in the pseudogap state.
\section{Case of disordered  normal-metal contact}
\label{sec:Cooperon}
\subsection{General considerations}
\label{sec:dirty_intro}
In Sec.~\ref{sec:clean} we investigated the conductance of a 
mesoscopic normal metal--to--pseudogap junction for the case of 
a perfectly clean normal metal.  We now address the issue of the 
sensitivity of the main result  (i.e.~that this conductance 
contains information regarding the spatial extent of the 
pseudogap-side phase correlations) to the assumption that the 
contact is a perfectly clean metal~\cite{Ref:vanWees}.  
Specifically, we examine  how Eq.~(\ref{eq:fin}) is modified by the 
presence of disorder in the normal-metal contact.  As we shall see, 
in the presence of disorder the most significant contribution to 
the conductance is associated with so-called Cooperon diagrams, 
familiar from the theory of the weak-localization corrections to 
the conductivity of a disordered metal~\cite{REF:Rammer}.

As is conventional, we take the disorder to be due to uncorrelated 
point-like impurities, which scatter the electrons elastically. 
Moreover, we assume that the dephasing length $L_{\phi}$ is long, compared 
to both $\xf$ and the mean free path $\ell$ (which characterizes the 
strength of the disorder and is related to the scattering time $\tau$ via 
$\ell\equiv v_{\rm F}\tau$).  Although we are 
focusing on situations in which $L_{\phi}$
is larger than the interface size~\cite{ref:temp},
 so that one expects substantial sample-to-sample 
fluctuations (which may in fact be interesting to study), we shall restrict 
our attention solely to the disorder average of the current.  
Then, averaging the current in Eq.~(\ref{final})
over configurations of the potential scatterers on the 
normal side, an averaging that we indicate via 
$\langle\cdots\rangle_{\rm dis}$, we arrive at
\begin{eqnarray}
\langle &&I(V)      
\rangle_{\rm dis} =
\frac{e}{\hbar}\frac{\pi}{8}|\tun_0 a|^4
     \tilde{\nu}_{\rm P}^2 
\int_{\curA} d^2\rho_1 \, \int_{\curA}d^2\rho_2 \,  
   g({\bbox{\rho}}_1-{\bbox{\rho}}_2)
\nonumber \\
&&\qquad
\times \int_{-\mu}^{2eV+\mu} d\epsilon \, 
\frac{\Delta^2}{\Delta^2-(eV-\epsilon)^2}
\big\{n(\epsilon-2eV)-n(\epsilon)\big\}
\nonumber \\
&& \qquad
\times 
\langle A^{\rm D}({\bbox{\rho}}_1,{\bbox{\rho}}_2;\epsilon) \,
A^{\rm D}({\bbox{\rho}}_1,{\bbox{\rho}}_2;2eV-\epsilon)
\rangle_{\rm dis}\,\,,
\label{disorderfinal}
\end{eqnarray}
where the superscript D refers to the disordered case.
The disorder-averaged product of spectral functions 
$\langle A^{\rm D}({\bbox{\rho}}_1,{\bbox{\rho}}_2;\epsilon) \,
A^{\rm D}({\bbox{\rho}}_1,{\bbox{\rho}}_2;2eV-\epsilon)
\rangle_{\rm dis}$
contains contributions that extend over length-scales 
$|{\bbox{\rho}}_1 -{\bbox{\rho}}_2|$
much larger 
than $\ell$.  These Cooperon contributions
provide the mechanism for the transmission of  the phase-sensitive 
information that would be probed in Andreev interferometry experiments 
involving a disordered normal-metal contact. 
\subsection{Semiclassical picture}
\label{sec:dirty_physics}
Before deriving our result for the contribution of the Cooperon to 
$\langle I(V)      
\rangle_{\rm dis}$, we pause to motivate physically why this 
particular
contribution is significant.  In the context of weak 
localization, the Cooperon contribution to the conductance
is usefully pictured in terms 
of constructive interference of pairs of paths involving the 
scattering of electrons from impurities in reverse 
order~\cite{REF:Larkin}.  This interference tends to ``localize'' 
electrons, thus causing a reduction in conductivity.  In the 
present context, however, the origin of the Cooperon is slightly 
different.  
To see this, consider the amplitude $A_{{\bf r}{\bf x}}^{\bar{\beta}\alpha}$ 
for an electron at position ${\bf x}$ in the normal metal to 
scatter from a sequence of impurities labeled by the index $\alpha$, 
to Andreev reflect at the position ${\bf r}$ on the interface, to then 
scatter from the sequence of impurities labeled $\bar{\beta}$, 
and finally to return to the position ${\bf x}$.  Then, the 
probability for an electron leaving ${\bf x}$ and reflecting from 
the interface to return to ${\bf x}$ as a hole is given by the 
squared modulus of the sum of such amplitudes, i.e., 
\begin{equation}
P({\bf x}) =
\Big|\sum_{\alpha\bar{\beta}{\bf r}}
A_{{\bf r}{\bf x}}^{\bar{\beta}\alpha}\Big|^2 
=
\sum_{\alpha\bar{\beta}{\bf r}}
\sum_{\alpha'\bar{\beta}'{\bf r}'}
\big(A_{{\bf r} {\bf x}}^{\bar{\beta}\alpha}\big)^{*}\, 
     A_{{\bf r}'{\bf x}}^{\bar{\beta}'\alpha'}.
\label{eq:4.2b}
\end{equation}
\begin{figure}[hbt]
\epsfxsize=6.0cm
 \centerline{\epsfbox{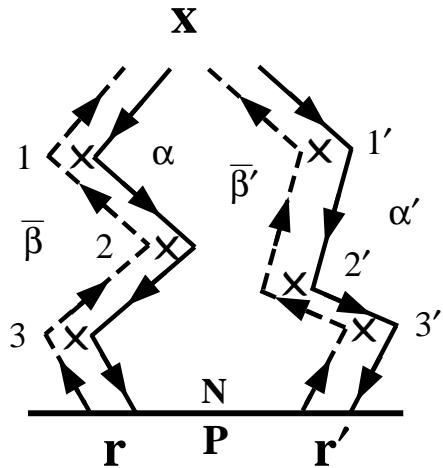}}
\vskip+.2truecm 
\caption{Schematic depiction of two semiclassical trajectories in 
which an electron leaves position ${\bf x}$ in the normal region 
(N), undergoes multiple elastic scattering events, then undergoes 
Andreev reflection at the NS interface (horizontal line), and then 
returns to ${\bf x}$ as a hole  via the same scatterers but in 
reverse order. Full (dashed) lines represent electron (hole) 
trajectories; crosses represent impurity scattering potentials.} 
\label{FIG:schematic} 
\end{figure}
As is well known, the amplitude $A_{{\bf r}{\bf x}}^{\bar{\beta}\alpha}$  
depends sensitively on the specific realization of the disorder; thus,
the right-hand side of  Eq.~(\ref{eq:4.2b}) contains many terms that
are disorder-dependent complex numbers.  These contributions to 
$P({\bf x})$ average to zero upon disorder averaging.  However, 
amongst the collection of amplitudes there is a special subset 
describing processes in which the hole, as it returns from the 
interface to ${\bf x}$, does so via the same collection of impurities 
visited by the electron on the outbound segment of the trajectory 
but in the reverse order.  If we denote the reverse of the sequence of 
impurities $\bar{\beta}$ by the sequence $\beta$ then this special 
subset consists of the amplitudes $A_{{\bf r}{\bf x}}^{\alpha\alpha}$;  
these amplitudes have the form of {\it real\/} numbers, regardless of 
the specific locations of the impurities, except for a factor 
due to the phase shift associated with Andreev reflection.  To 
see this, consider, e.g.,  the left-hand pair of paths (electron and 
hole) in Fig.~\ref{FIG:schematic}.  The electron path (full line) 
originates at ${\bf x}$, scatters from impurities at positions $1$, $2$ 
and  $3$, and then Andreev reflects as a  hole.  The hole then propagates 
back to  ${\bf x}$, scattering from the impurities at positions
$3$, $2$ and $1$ before returning to the position ${\bf x}$.  
The dynamical phase acquired by the electron as it propagates
to the interface is canceled by a phase of the opposite
sign acquired by the hole.  
Thus, the amplitude for an electron at ${\bf x}$ to return as a 
hole at  ${\bf x}$ depends only on the phase of the condensate at 
${\bf r}$:
\begin{equation}
A_{{\bf r}{\bf x}}^{\alpha \alpha} 
\sim {\rm e}^{i\phi({\bf r})}. 
\end{equation}
Thus, one sees that the most significant contribution to 
Eq.~(\ref{eq:4.2b}) is given approximately by
\begin{equation}
P({\bf x}) \sim \sum_{{\bf r},{\bf r}'} 
{\rm e}^{i(\phi({\bf r})-\phi({\bf r}'))},
\end{equation}
and hence that $P({\bf x})$ is sensitive to the
nature of the pseudogap phase-phase correlations,
a sensitivity similar to that embodied in Eq.~(\ref{eq:fin}).

\subsection{Microscopic calculation}
The explicit computation of the contribution of the Cooperon directly 
follows the usual analysis found in the context of weak localization; 
following Rammer~\cite{REF:Rammer},
we find that the disorder-averaged product of spectral functions
has the form
\begin{eqnarray}
&& \left\langle
 A^{\rm D}({\bf x},{\bf x}',\epsilon)
 A^{\rm D}({\bf x},{\bf x}',\epsilon')\right\rangle_{\rm dis} 
\label{eq:spectralprod}
\\
&&= \frac{2\pi}{\hbar}  \nu_{\rm N}\!\int\! \frac{d^3 Q}{(2\pi\hbar)^3} 
  {\rm e}^{i{\bf Q}\cdot ({\bf x}-{\bf x}')/\hbar}
  \left\{C({\bf Q},\epsilon-\epsilon')\! +\! 
C({\bf Q},\epsilon'-\epsilon)\right\},
\nonumber
\end{eqnarray}
where the Cooperon propagator
$C({\bf Q},w) \equiv (-i\omega/\hbar + DQ^2\hbar^{-2})^{-1}$,
the diffusion constant $D \equiv v_{\rm F}^2 \tau/3$ (in
three dimensions), and  $\nu_{\rm N} [\equiv \kf m/(2\pi^2 \hbar^2)]$
is the normal-side density of states. 
Inserting Eq.~(\ref{eq:spectralprod}) into Eq.~(\ref{disorderfinal}) leads to the expression
\begin{eqnarray}
&&\langle I(V)\rangle_{\rm dis}
=
e \left|\frac{\tun_0\kf a}{4\pi\ef}\right|^4
\frac{\pi^2 \kf^3}{2m}
                  \int_{\curA} d^2 {\rho}_1
\int_{\curA}d^2 {\rho}_2 \,\,
                  g({\bbox{\rho}}_1-{\bbox{\rho}}_2)
\nonumber
\\
&&\times\int_{-\mu_{\rm N}}^{2eV+\mu_{\rm N}} d\epsilon\,
      \big\{n(\epsilon-2eV) - n(\epsilon)\big\} 
\frac{\Delta^2}{\Delta^2 - (eV-\epsilon)^2}
\nonumber \\
&&\times 
\big\{C\big({\bbox{\rho}}_1-{\bbox{\rho}}_2, 2(eV-\epsilon)\big) +
 C\big({\bbox{\rho}}_1-{\bbox{\rho}}_2, 2(\epsilon-eV)\big)\big\},
 \label{eq:findis1}
\end{eqnarray}
where  $C\big({\bbox{\rho}}, \epsilon\big)$ is the (three-dimensional) Fourier 
transform of $C\big({\bf Q},\epsilon\big)$.

To analyze $\langle I(V)\rangle_{\rm dis}$ we 
make two further simplifying assumptions.  
First, as in the clean case, we limit our attention to 
low temperatures (i.e.~$\kb T \ll eV$), and thus we obtain
\begin{eqnarray}
&&\langle I(V)\rangle_{\rm dis} \approx
\frac{e}{\hbar} \left|\frac{\tun_0\kf a}{4\pi\ef}\right|^4
 \frac{\pi\hbar}{4mD} \kf^3
\int_{0}^{2eV} d\epsilon\, 
\frac{\Delta^2}{\Delta^2 - (eV-\epsilon)^2}
\nonumber
\\ 
&& \quad\times
                  \int_{\curA} d^2 {\rho}_1 \,
                  \int_{\curA} 
d^2 {\rho}_2 \,\,
                  \frac{g({\bbox{\rho}}_1-{\bbox{\rho}}_2)}
                    {|{\bbox{\rho}}_1-{\bbox{\rho}}_2|}  
{\rm e}^{-\sqrt{(\hbar D)^{-1}|eV-\epsilon|} 
        |{\bbox{\rho}}_1-{\bbox{\rho}}_2|} 
\nonumber \\
&&\quad\times 
\cos\big\{\sqrt{(\hbar D)^{-1}|eV-\epsilon|}
        |{\bbox{\rho}}_1-{\bbox{\rho}}_2|\big\},
\label{eq:disorder2}
\end{eqnarray}  
where we have inserted the explicit real-space expression for the 
Cooperon.  Second, by making the restriction to low voltages
(i.e.~$eV \ll \Delta$)~\cite{REF:inequals}, 
we may in Eq.~(\ref{eq:disorder2}) replace 
$\Delta^2/(\Delta^2 - (eV-\epsilon)^2)$ by $1$.  Furthermore, 
in the presence of disorder one has the natural length-scale 
$L_V \equiv \sqrt{\hbar D/eV}$.  
At sufficiently low voltages and small interface 
sizes, $L_V$ will be much larger than typical values of  
$|{\bbox{\rho}}_1-{\bbox{\rho}}_2|$,
so that one may expand to lowest order in  $L/ L_V$, thus obtaining
\begin{eqnarray}
\label{eq:disorder3}
&&\langle I(V)\rangle_{\rm dis}
\nonumber \\
&&
\qquad \approx
\nonumber
\frac{e^2}{\hbar}V \left|\frac{\tun_0\kf a}{4\pi\ef}\right|^4
 \frac{\pi \kf}{L_{\rm F}^2 L_V} 
                  \int_{\curA} d^2 {\rho}_1
                  \int_{\curA} d^2 {\rho}_2 \,\,
                  g({\bbox{\rho}}_1-{\bbox{\rho}}_2)
\\
&&\qquad 
\Big\{
\frac{L_V}{|{\bbox{\rho}}_1-{\bbox{\rho}}_2|}  
+ {\cal O}(1) \Big\},
\label{eq:findis}
\end{eqnarray}
where $L_{\rm F} \equiv \sqrt{\hbar D/\ef}$.
Thus, as  found  in Sec.~\ref{sec:clean} for the case of a clean 
normal metal contact, the low-voltage conductance of a disordered metal--to--pseudogap junction also contains information regarding 
the pseudogap phase-phase correlation function.  
\subsection{Illustrative example: BKT correlations}
\label{SEC:illu_BKT}
In this section we examine the area-dependence of the 
low-temperature and 
low-voltage conductance of a disordered normal 
metal--to--pseudogap junction for the case of BKT correlations. 
As in Sec.~\ref{sec:clean}, we assume
that phase correlations decay in an exponential fashion, consistent with
the BKT scenario.
Our starting point is thus Eq.~(\ref{eq:findis}), together with 
the model of phase correlations given 
by Eq.~(\ref{eq:g(r)}).  
By considering the  $V \rightarrow 0^{+}$ behavior of
 Eq.~(\ref{eq:findis}) we arrive at the low-temperature
conductance per unit area (for the case of an interface having the 
shape of a disk of radius $L$):
\begin{mathletters}
\begin{eqnarray}
&&\left.
\frac{\langle I(V)\rangle_{\rm dis}}{\pi L^2 \,V}
\right|_{V\rightarrow 0^{+}} 
\approx
\Gamma_{\rm D}\, 
f_{\rm D}(L/\xf),
\\
&&\Gamma_{\rm D} \equiv  
\frac{e^2}{\hbar} \left|\frac{\tun_0\kf a}{4\pi\ef}\right|^4
   \frac{2\pi^2 \kf \xf}{L_{\rm F}^2},
\\
&&f_{\rm D} \equiv 
\frac{1}{2\pi^2} \Big(\frac{L}{ \xf} 
\Big)
\int_{1} d^2x_1  \int_{1} d^2x_2   
          \frac{{\rm e}^{-|{\bf x}_1-
                    {\bf x}_2|(L/\xf)} }
{|{\bf x}_1-{\bf x}_2|}. \label{eq:4.8a}
\end{eqnarray}
\end{mathletters}
\begin{figure}[hbt]
\epsfxsize=8.0cm
 \centerline{\epsfbox{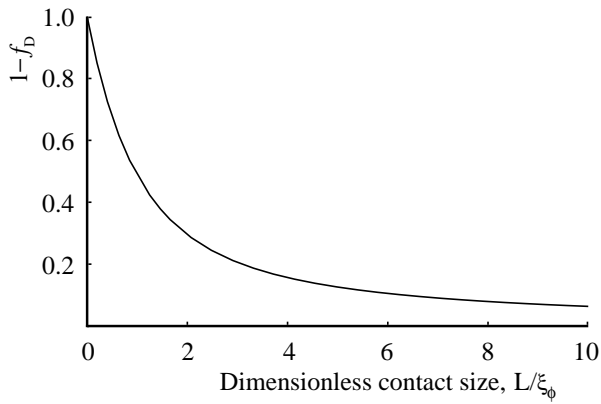}}
\vskip+.2truecm 
\caption{$1-f_{\rm D}$ (i.e.~the departure of the 
dimensionless Andreev conductance per unit area from its 
large-area limit) as a function of the dimensionless size 
of the interface $L/\xf$ for the case of a disordered 
normal-metal contact (computed numerically).  For $L$ much 
smaller than $\xf$, the zero-voltage conductance per unit 
area is much smaller than its asymptotic value.} 
\label{FIG:fdis} 
\end{figure}
 
By evaluating the integrals in Eq.~(\ref{eq:4.8a}), we obtain 
\begin{equation}
f_{\rm D}= 1+ \left(\frac{\xf}{L}\right)\,\left\{
L_1\left(\frac{2L}{\xf}\right)- I_1\left(\frac{2L}{\xf}\right) 
\right\}
\end{equation}
where $I_1$ is a modified Bessel function and $L_1$ is a modified
Struve function.  The asymptotic behavior of $f_{\rm D}$ is linear 
for small $L$ 
(i.e.~$f_{\rm D}\approx (8/3\pi)(L/\xf)$ for $ L/\xf \ll 1$);
for large $L$ it approaches unity as an inverse power of $L$
[i.e.~$f_{\rm D} \approx 1 - (2\xf/\pi L)$ for $L/\xf \gg 1$].
In Fig.~\ref{FIG:fdis} we show how this function crosses over between these
two limits.  
As with the case of the clean contact, the conductance shows marked 
sensitivity to the phase-phase correlations of the 
pseudogap state.
\section{Concluding remarks}
In this Paper, we have proposed and explored theoretically 
the possibility of using Andreev interferometry to probe the 
spatial structure of the phase correlations in the pseudogap 
state of the cuprate superconductors.  The viability of this 
technique rests on the sensitivity of the tunneling current 
across mesoscopic
 normal--to--pseudogap junctions to spatial variations 
in the local superconducting order in the pseudogap state, as 
well as the possibility of using junctions having a range of 
areas.

By considering a simple tunneling model, we have established 
a relationship between the tunneling current and the 
equilibrium phase-phase correlator characterizing the 
pseudogap state.  We have considered the cases in which the 
normal region (i.e.~the contact) is either a clean or a 
disordered metal.  In both cases, we have assumed that phase 
coherence length for quasiparticles on the normal side is 
greater than the contact size.  If this 
condition is not met then, 
is not the case then, 
throughout our results, the interface size would have to  be replaced 
by the dephasing length. 

To illustrate this Andreev interferometry proposal, we have 
applied our general results to a simple model of the pseudogap 
phase-phase correlations, which is intended to mimic the BKT 
correlations relevant to certain proposed pictures of the 
pseudogap state.  Our considerations suggest that measurements 
of the low-voltage conductance of mesoscopic tunnel junctions of 
varying areas between normal-state and pseudogap-state regions 
would reveal information about the phase-phase correlations in 
the pseudogap state. 
\noindent
\section*{Acknowledgments}
\label{SEC:Acknowledgments}
It is a pleasure to thank Yuli Lyanda-Geller, David Pines
 and Alexander Shnirman 
for extensive discussions.  This work was supported by the 
U.S.~Department of Energy, Division of Materials Sciences, 
under Award No.~DEFG02-96ER45439, through the University of 
Illinois Materials Research Laboratory (D.E.S., P.M.G.),
the Science and Technology Center for 
Superconductivity through NSF grant DMR 91-20000 (J.S.), 
the Deutsche Forschungsgemeinschaft
(J.S.), and 
the NSF via grant DMR98-75565 (A.Y.). 

\end{multicols}

\begin{references}
\bibitem[\ast]{REF:DES} 
Electronic address: {\tt d-sheehy@uiuc.edu\/}
\bibitem[\dagger]{REF:PMG} 
Electronic address: {\tt
goldbart@uiuc.edu\/}
\bibitem[\ddagger]{REF:JS} 
Current address: 
Ames Laboratory, 
Iowa State University, 
1 Osborn Drive, 
Ames, IA~50011; 
electronic address: 
{\tt schmal@cyclops.ameslab.gov\/}
\bibitem[{\mathchar "278}]{REF:AY} 
Electronic address: {\tt ayazdani@uiuc.edu\/}
\bibitem{AGL96} 
A. G. Loeser, Z.-X. Shen, D. S. Dessau, D. S. Marshall, C. H. Park, 
P. Fournier and  A. Kapitulnik, 
Science {\bf 273\/}, 325 (1996).
\bibitem{HD96}
H. Ding, T. Yokoya, J. C. Campuzano, T. Takahashi, M. Randeria,
M. R. Norman, T. Mochiku, K. Kadowaki and J. Giapintzakis,  
Nature (London) {\bf 382\/}, 51 (1996).
\bibitem{CROF98}
Ch. Renner, B. Revaz, J.-Y. Genoud, K. Kadowaki and {\O}. Fischer, 
Phys. Rev. Lett. {\bf 80\/}, 149 (1998).
\bibitem{LN92}
P. A. Lee and N. Nagaosa, 
\prb {\bf 45}, 966 (1992);
{\rm ibid.} {\bf 46}, 5621 (1992).
\bibitem{PWA97} P. W. Anderson, {\sl Theory of Superconductivity in
the High-T$_c$ Cuprate Superconductors\/} (Princeton, NJ, 1997).
\bibitem{SPS98} 
J. Schmalian, D. Pines and B. Stojkovic,
Phys. Rev. Lett. {\bf 80}, 3839 (1998).
\bibitem{EK95} 
V. Emery and S. A. Kivelson, 
Nature (London) {\bf 374\/}, 434 (1995).
\bibitem{ML96} J. Maly and K. Levin, 
Phys. Rev. B {\bf 54}, R15657
(1996); 
B. Jank\'o, J. Maly and K. Levin, 
Phys. Rev. B {\bf 56}, R11407
(1997).
\bibitem{ER97} J. Engelbrecht, M. Randeria and C. A. R. S\'a de Melo,
Phys. Rev. B {\bf 55\/}, 15153 (1997).
\bibitem{SGB97} J. Schmalian, S. Grabowski and K. H. Bennemann,
Phys. Rev. B {\bf 56\/}, R509 (1997).
\bibitem{CAV98} A. V. Chubukov, 
Europhys. Lett. {\bf 44}, 655 (1998).
\bibitem{Janko} 
A scheme for probing the energetic structure of pairing fluctuations in 
the pseudogap regime has recently been proposed and developed 
by Jank\'o and collaborators; see Refs.~\cite{Janko2,Janko3}.
\bibitem{Janko2}
B. Jank\'o, I. Kosztin, K. Levin, M. R. Norman and D. J. Scalapino, 
Phys. Rev. Lett. {\bf 82}, 4304 (1999).
\bibitem{Janko3}
B. Jank\'o (unpublished, 1998) has extended the approach of 
Ref.~\cite{Janko2} to the cases of pseudogap-pseudogap and 
normal-pseudogap tunneling.
\bibitem{REF:avoid} 
We shall assume that the adopted  tunneling geometry
minimizes cancellations in the tunneling amplitude due
to the d-wave form of the pairing state.  The phase referred to in the
present Paper is the overall phase of the order parameter.
\bibitem{FM98} M. Franz and A. J. Millis,
Phys. Rev. B {\bf 58\/}, 14572 (1998).
\bibitem{KD98} 
H.-J. Kwon and  A. T. Dorsey, 
Phys. Rev. B {\bf 59}, 6438 (1999).
\bibitem{REF:conduct} 
See, e.g.,
Refs.~\cite{REF:Andreev,REF:Griffin,REF:BTK}.
\bibitem{REF:Andreev} 
A. F. Andreev,  
Zh. Eksp. Teor. Fiz {\bf 46}, 1823 (1964) 
[Sov. Phys. JETP {\bf 19}, 1228 (1964)].
\bibitem{REF:Griffin} 
J. Demers and A. Griffin, 
Can. J. Phys. {\bf 49\/}, 285 (1970); 
A. Griffin and J. Demers, 
Phys. Rev. B {\bf 4\/}, 2202 (1971).
\bibitem{REF:BTK} G. E. Blonder, M. Tinkham and T. M. Klapwijk,
Phys. Rev. B {\bf 25\/}, 4515 (1982).
\bibitem{REF:Spivak} 
See, e.g, B. Z. Spivak and
D. E. Khmel'nitski\u\i, 
Pis'ma Zh. Eksp. Teor. Fiz. {\bf 35}, 334 (1982) 
[JETP Lett. {\bf 35}, 412 (1982)].
\bibitem{REF:Hekking} 
F. W. J. Hekking and Yu. V. Nazarov,
Phys. Rev. Lett. {\bf 71\/}, 1625 (1993).
\bibitem{REF:Andint} See, e.g., 
 A. Dimoulas, J. P. Heida, B. J. v. Wees, 
 T. M. Klapwijk, W. v. d. Graaf, and G. Borghs,
 Phys. Rev. Lett. {\bf 74\/}, 602 (1995);
P. G. N. de Vegvar, T. A. Fulton, W. H. Mallison, and R. E. Miller, 
Phys. Rev. Lett. {\bf 73\/}, 1416 (1994);
H. Pothier, S. Gu\'eron, D. Esteve, and M. H. Devoret,  
Phys. Rev. Lett. {\bf 73\/}, 2488 (1994); 
H. Nakano and H. Takayanagi, 
Phys. Rev. B {\bf 47\/}, 7986 (1993);
C. J. Lambert, 
J. Phys. Condens. Mat. {\bf 5}, 707 (1993).
\bibitem{REF:Choi} 
H.-Y. Choi, Y. Bang and D. K. Campbell,
cond-mat/9902125.
\bibitem{REF:Covington}
M. Covington, R. Scheuerer, K. Bloom and L. H. Greene
Appl. Phys. Lett.  {\bf 68\/}, 1717 (1996).
\bibitem{Mahan} For a review of the tunneling formalism, see, e.g.,
G. D.  Mahan, {\it Many Particle Physics\/} 
(Plenum, New York, 1990), Sec.~9.3.
\bibitem{Tsuzuki} 
As applied, e.g., by 
T. Tsuzuki, 
Prog. Theor. Phys. {\bf 41\/}, 1600 (1969).
\bibitem{ref:corson} 
J. Corson, R. Mallozzi, J. Orenstein, J. N. Eckstein and I. Bozovic, 
Nature (London) {\bf 398\/}, 221 (1999). 
\bibitem{REF:AGD} See, e.g., A. A. Abrikosov, L. P. Gorkov, and
I. E. Dzyaloshinski, {\it Methods of Quantum Field Theory in
Statistical Physics\/} (Dover, New York, 1975), Chap. 7.
\bibitem{d-wave}
To treat the case of d-wave pairing, one would need to take into 
account directional anisotropy in the tunneling matrix element
as well as in $\Delta_{\bf k}$.  We expect that the results 
will be qualitatively the same for this case. 
\bibitem{REF:inequals} 
We note that the assumptions leading to Eq.~(\ref{eq:fin}) or 
(\ref{eq:findis}) (i.e.~$\kb T \ll eV$ and $eV \ll \Delta$) imply 
that $T$ must be chosen to be much smaller than $\Delta$.  
Thus, we expect the experiment to be most feasible for strongly 
underdoped cuprates, in which $\tc$ is small relative to $\Delta$.
\bibitem{Ber70} 
L. Berezinski\i, 
Zh. Eksp. Teor. Fiz {\bf 59}, 907 (1970) 
[Sov. Phys. JETP {\bf 32}, 493-500 (1971)].
%
\bibitem{KT73} 
J. M. Kosterlitz and  D. J. Thouless, 
J. Phys. C {\bf 6\/}, 1181 (1973).
\bibitem{Kosterlitz}
J. M. Kosterlitz, 
J. Phys. C {\bf 7\/}, 1046 (1974).
\bibitem{REF:exEK95}
See, e.g., Ref.~\cite{EK95}. 
\bibitem{Ref:vanWees}
The interplay between Andreev reflection and disorder 
on the normal side of an NS junction was already discussed
in the literature some time ago; 
see
B. J. van Wees, P. de Vries, P. Magn\'ee, and T. M. Klapwijk,
Phys. Rev. Lett. {\bf 69}, 510 (1992) and 
 Ref.~\cite{REF:Hekking} .
\bibitem{REF:Rammer}
For a pedagogical review of the effects of disorder in 
conductors, see, e.g., 
J. Rammer, 
{\sl Quantum Transport Theory}
(Perseus, Reading, 1998).
\bibitem{ref:temp}
For the dephasing length in the normal metal contact to be 
long compared with the interface size
we expect that the system will have to be at
low temperatures.  Higher temperatures will also smear the 
current-voltage signal 
[cf. the Fermi distribution in
Eq.~(\ref{final})] 
Thus, we expect
that the experiment will necessarily involve strongly 
underdoped cuprate materials with a small $\tc$.
\bibitem{REF:Larkin}
 See, e.g., A. I. Larkin and D. E. Khmel'nitski\u\i, 
Usp. Fiz. Nauk {\bf 136}, 533 (1982) [Sov. Phys. Usp. {\bf 25}, 185
(1982)]; G. Bergmann, Phys. Rep. {\bf 107}, 1 (1984); 
S. Chakravarty and A. Schmid, Phys. Rep. {\bf 140}, 193 (1986).   
\end{references}
\end{document}